\documentclass[final]{siamltex}

\def\QEDclosed{\mbox{\rule[0pt]{1.3ex}{1.3ex}}} % for a filled box

\usepackage{epsf}
\usepackage{amsmath}
\usepackage{amscd}
\usepackage{amssymb}
\usepackage{graphicx}
\usepackage{stmaryrd}

\begin{document}

\newtheorem{thm}{Theorem}
\newtheorem{example}{Example}
\newtheorem{algorithm}{Algorithm}

% paper title
\title{Separable Convex Optimization Problems with Linear Ascending Constraints
\thanks{This work was supported by the University Grants Commission
under Grant Part (2B) UGC-CAS-(Ph.IV) and by the Department of Science and Technology under Grant DSTO748.}}

\author {Arun~Padakandla
\and
Rajesh~Sundaresan\thanks{Department of Electrical Communication
Engineering, Indian Institute of Science, Bangalore 560012, India. E-mail:~{\tt rajeshs@ece.iisc.ernet.in}}}

% make the title area
\maketitle

\begin{picture}(0,0)
  \put(0,100){\tt\small Submitted to the SIAM Journal on Optimization, Jul. 2007.}
\end{picture}

\begin{abstract}
Separable convex optimization problems with linear ascending inequality and equality constraints are addressed in this paper. Under an ordering condition on the slopes of the functions at the origin, an algorithm that determines the optimum point in a finite number of steps is described. The optimum value is shown to be monotone with respect to a partial order on the constraint parameters. Moreover, the optimum value is convex with respect to these parameters. Examples motivated by optimizations for communication systems are used to illustrate the algorithm.
\end{abstract}



\pagestyle{myheadings}
\thispagestyle{plain}
\markboth{A. PADAKANDLA AND R. SUNDARESAN}{CONVEX OPTIMIZATION WITH ASCENDING CONSTRAINTS}

\section{Problem description}
\label{sec:problemDescription}

Let $g_m, m=1,2,\cdots,L$ be functions that satisfy the following:
\begin{itemize}
\item $g_m : (a_m,b_m) \rightarrow \mathbb{R}$ where $a_m \in [-\infty, 0)$ and $b_m \in (0, +\infty]$ and therefore $a_m < 0 < b_m$;
\item $g_m$ is strictly convex in its domain $(a_m, b_m)$;
\item $g_m$ is continuously differentiable in its domain $(a_m, b_m)$;
\item The slopes of the functions at 0, i.e., the values of the strictly increasing function $h_m := g_m'$ at 0, are in increasing order with respect to the index $m$:
\begin{equation}
  \label{eqn:OrderingDerivativeAt0}
  h_1(0) \leq h_2(0) \leq \cdots \leq h_L(0);
\end{equation}
\item There is a point in the domain $(a_m, b_m)$ where the slope of $g_m$ equals $h_1(0)$, the slope of the first function at 0. (This may be equivalently stated as $h_1(0) \geq h_m(a_m+)$, given (\ref{eqn:OrderingDerivativeAt0}) and that $h_m$ is continuous and strictly increasing).
\end{itemize}
In this paper, we minimize the separable objective function $G:\mathbb{R}^L \rightarrow \mathbb{R}$ given by
\begin{equation}
  \label{eqn:minimizeSumSeparableConvexFunctions}
  G(y) := \sum_{m=1}^L g_m(y_m),
\end{equation}
where $y = (y_1, \cdots, y_L)$, subject to the following linear inequality and equality constraints:
\begin{eqnarray}
  \label{eqn:positivityBounded}
  y_m & \in & [0, \beta_m], ~~m=1,2,\cdots,L,\\
  \label{eqn:ladderConstraint}
  \sum_{m=1}^{l} y_m & \geq & \sum_{m=1}^{l} \alpha_m, ~~l=1,2,\cdots,L-1, \\
  \label{eqn:equalityConstraint}
  \sum_{m=1}^{L} y_m & =    & \sum_{m=1}^{K} \alpha_m.
\end{eqnarray}
In the above constraints, we assume $\beta_m \in (0, b_m]$ for $m = 1, 2, \cdots, L$, $\alpha_m \geq 0$ for $m = 1, 2, \cdots, K$, where $K \geq L$, and naturally,
\begin{equation}
  \label{eqn:commonSense1}
  \sum_{m=1}^K \alpha_m \leq \sum_{m=1}^L \beta_m.
\end{equation}
We also assume
\begin{equation}
  \label{eqn:commonSense2}
  \sum_{m=L}^K \alpha_m > 0.
\end{equation}
The inequalities in (\ref{eqn:positivityBounded}) impose positivity and upper bound constraints. Note that if $\beta_m = b_m$, the upper bound constraint is irrelevant because the domain of $g_m$ is $(a_m,b_m)$. The inequalities in (\ref{eqn:ladderConstraint}) impose a sequence of {\it ascending constraints} with increasing heights $\sum_{m=1}^l \alpha_m$ indexed by $l$. Assumption (\ref{eqn:commonSense1}) is necessary for the constraint set to be nonempty. Without (\ref{eqn:commonSense2}), it is easy to see that $y_L = 0$, and the problem reduces to a similar one with fewer variables.

What we have described is a separable convex optimization problem with linear inequality and equality constraints. A rich duality theory exists for such problems. See Bertsekas \cite[Sec. 5.1.6]{2003xxNLP_Ber}. Here, we provide an algorithm that puts out a vector that minimizes (\ref{eqn:minimizeSumSeparableConvexFunctions}) and terminates in at most $L$ steps. Section \ref{sec:mainResults} contains a description of the algorithm and Section \ref{sec:proofs} the proof of its optimality. While we may take $K = L$ without loss of generality, allowing $K \geq L$ simplifies the exposition of our algorithm.

Problems of the above kind arise in the optimization of multi-terminal communication systems where power utilized, measured in Joules per second, is minimized, or throughput achieved, measured in bits per second, is maximized, subject to meeting certain quality of service and feasibility constraints. See Viswanath \& Anantharam \cite{200206TIT_VisAna} for details and Section \ref{sec:examples} for specific examples. Viswanath \& Anantharam \cite{200206TIT_VisAna} provide two algorithms for their power minimization and throughput maximization problems. Our work unites their solutions and goes further to minimize any $G$ that satisfies the constraints mentioned above. Under a further condition on the functions which will be stated in Section \ref{sec:mainResults}, we argue that our algorithm provides the solution to the above optimization problem with the additional ordering constraint $y_1 \geq y_2 \geq \cdots \geq y_L$.

\section{The Main Results}
\label{sec:mainResults}

We begin with some remarks on notation.

\begin{itemize}
\item For integers $i,j$ satisfying $i \leq j$, we let $\llbracket  i,j  \rrbracket $ denote the set $\{ i, i+1, \cdots, j \}$.

\item Let $\mathbb{E}_m := h_m(~(a_m, b_m)~)$, the range of $h_m$. Thus the condition $h_1(0) > h_m(a_m+), m \in \llbracket 1,L  \rrbracket $ in Section \ref{sec:problemDescription} may be written as
    \begin{equation}
      \label{eqn:h10Condition}
      h_1(0) \in \cap_{m=1}^L \mathbb{E}_m.
    \end{equation}

\item Denote by $h_m^{-1}: \mathbb{E}_m \rightarrow (a_m, b_m)$ the inverse of the continuous and strictly increasing function $h_m$. The inverse is also continuous and strictly increasing in its domain.

\item For convenience, define the functions $H_m: \mathbb{E}_m \rightarrow (a_m, \beta_m]$ to be
  \begin{equation}
    \label{eqn:H}
    H_m := h_m^{-1} \wedge \beta_m.
  \end{equation}
  $H_m$ is clearly increasing.\footnote{We say $f$ is increasing if $a>b$ implies $f(a) \geq f(b)$. If there is strict inequality, we say $f$ is strictly increasing. Similarly we use positive for $\geq 0$ and strictly positive for $>0$.} Assignments to the variable $y_m$ will be via evaluation of $H_m$ so that the upper bound constraint in (\ref{eqn:positivityBounded}) is automatically satisfied.

\item For $1 \leq i \leq l < L$, let $\theta_i^l$ denote the least $\theta \geq h_1(0)$ that satisfies the equation
  \begin{equation}
    \label{eqn:smallTheta}
    \sum_{m=i}^l H_m(\theta) = \sum_{m=i}^l \alpha_m,
  \end{equation}
  provided the set of such $\theta$ is nonempty. Otherwise we say $\theta_i^l$ does not exist. The domain of $\sum_{m=i}^l H_m$ is $\cap_{m=i}^l \mathbb{E}_m$. The function $\sum_{m=i}^l H_m$ is increasing, and moreover, strictly increasing until all functions in the sum saturate. So there is no solution to (\ref{eqn:smallTheta}) when for example $\sum_{m=i}^l \alpha_m > \sum_{m=i}^l \beta_m$. In general, if we can demonstrate the existence of $\underline{\theta}$ and $\overline{\theta}$, both in the set $\cap_{m=i}^l \mathbb{E}_m$, that satisfy
  \begin{equation}
    \label{eqn:underlineOverlineTheta}
    \sum_{m=i}^l H_m(\underline{\theta}) \leq \sum_{m=i}^l \alpha_m \leq \sum_{m=i}^l H_m \left(\overline{\theta}\right),
  \end{equation}
  then the existence of $\theta_i^l \in \cap_{m=i}^l \mathbb{E}_m$ is assured, thanks to the continuity of $\sum_{m=i}^l H_m$. Indeed, we may always take $\underline{\theta} = h_1(0)$. This is because our assumptions (\ref{eqn:h10Condition}), (\ref{eqn:OrderingDerivativeAt0}), and the increasing property of $H_m, m \in \llbracket 1,L  \rrbracket $ imply
  \[
    \sum_{m=i}^l H_m \left(h_1(0) \right) \leq \sum_{m=i}^l H_m \left( h_m(0) \right) = \sum_{m=i}^l \left( h_m^{-1} \left( h_m(0) \right) \wedge \beta_m \right) = 0 \leq \sum_{m=i}^l \alpha_m.
  \]
  The continuity and increasing property of $\sum_{m=i}^l H_m$ further imply that
  \begin{equation}
    \label{eqn:underlineOverlineThetaCondition}
    \left( \underline{\theta} \wedge h_1(0) \right) \leq \theta_i^l \leq \overline{\theta}.
  \end{equation}
  Thus, in order to show existence of $\theta_i^l$, it is sufficient to identify a $\overline{\theta}$ that satisfies the right side inequality of (\ref{eqn:underlineOverlineTheta}). We will have occasion to use this remark a few times in the proof of correctness of the algorithm.

  \item Similarly, for $1 \leq i \leq j \leq L$, we let $\Theta_i^j$ denote the least $\theta  \geq h_1(0)$ that satisfies the equation
  \begin{equation}
    \label{eqn:bigTheta}
    \sum_{m=i}^j H_m(\theta) = \sum_{m=i}^K \alpha_m,
  \end{equation}
  provided the set of such $\theta$ is nonempty. Otherwise we say $\Theta_i^j$ does not exist. The difference between (\ref{eqn:smallTheta}) and (\ref{eqn:bigTheta}) is the summation up to $K$ in the right side of (\ref{eqn:bigTheta}) and the consequent difference in the upper limits on the left and right sides of (\ref{eqn:bigTheta}). Hence the upper case $\Theta_i^j$. The remarks made above on the existence of $\theta_i^l$ are applicable to $\Theta_i^j$.

  \item We now provide a description of the variables used in the algorithm for ease of reference.
  \begin{itemize}
    \item $n$: Iteration number.
    \item $i_n$ and $j_n$: Pointer locations of the first and the last variables, $y_{i_n}$ and $y_{j_n}$, that are yet to be set.
    \item $N$ : The last iteration number in which a variable is set.
    \item $l,m$ : Temporary pointer locations that satisfy $l,m \in \llbracket i_n,j_n  \rrbracket $ in iteration $n$.
    \item $t$ : Pointer to the variable that satisfies the corresponding ascending constraint with equality; $t \in \llbracket i_n,j_n  \rrbracket $.
    \item $\xi_n$ : Choice of the best slope (marginal cost) in iteration $n$.
    \item $p_m$ : Iteration number when variable $y_m$ is set. (This is needed only in the proof).
    \item $c_m$ : A label that indicates the type of $\xi_n$ that set the variable $y_m$. The possible labels are $\{\mathcal{A}, \mathcal{A}^*\}$ for Step 3(a) of the algorithm, $\{\mathcal{B}^*\}$ for Step 3(b) of the algorithm, and $\{\mathcal{C}, \mathcal{C}^*\}$ for Step 3(c) of the algorithm. If $c_m$ is assigned an asterisked label, then the ascending constraint is met with equality for $y_m$. (This is needed only in the proof).
  \end{itemize}
\end{itemize}

We now provide a generalization of the two algorithms given by Viswanath \& Anantharam \cite{200206TIT_VisAna}.

\vspace*{.25cm}

\begin{algorithm}
\label{alg:universal}
{\tt
\begin{itemize}
  \item {\bf Inputs}: $K, L, (\alpha_1, \alpha_2, \cdots, \alpha_K), (\beta_1,\beta_2, \cdots, \beta_L)$.
  \item {\bf Output}: $y^* = \left(y_1^*,y_2^*,\cdots,y_L^*\right)$.
  \item {\bf Step 1: Initialization} Set $n \leftarrow 1, i_1 \leftarrow 1, j_1 \leftarrow L$ and go to {\bf Step 2}.
  \item {\bf Step 2: Termination} If $i_n > j_n$, then set $N \leftarrow n-1$, output the vector $y^* = \left(y_1^*,y_2^*, \cdots,y_L^*\right)$, and {\bf stop}. \\
  Else go to {\bf Step 3}.

\item {\bf Step 3}: Find $\Theta_{i_n}^{j_n}$, the solution of (\ref{eqn:bigTheta}) with $i \leftarrow i_n$ and $j \leftarrow j_n$. Also find $\theta_{i_n}^l$ for $l \in \llbracket  i_n, j_n - 1  \rrbracket $. These are solutions of (\ref{eqn:smallTheta}) for $i \leftarrow i_n$ and $l$ as chosen.\footnote{Theorem \ref{thm:correctnessOfAlgorithm} gives a sufficient condition when these quantities can be identified in every iteration.} Then set
\begin{equation*}
\label{IterationMax}
\xi_n = \max \left\{  \Theta_{i_n}^{j_n}, ~h_{j_n}(0), ~\theta_{i_n}^{l}; l \in \llbracket i_n, j_n-1  \rrbracket \right\}
\end{equation*}

\item {\bf Case 3(a)}: If $\xi_n = \Theta_{i_n}^{j_n}$, then set
\begin{eqnarray*}
y_m^* &\leftarrow& H_m \left(\xi_n \right) \mbox{ for } m \in \llbracket  i_n,j_n  \rrbracket  \nonumber\\
p_m &\leftarrow& n \mbox{ for } m \in \llbracket  i_n,j_n  \rrbracket  \nonumber\\
c_m &\leftarrow& \mathcal{A} \mbox{ for } m \in \llbracket  i_n,j_n-1  \rrbracket  \nonumber\\
c_{j_n} &\leftarrow& \mathcal{A}^* \nonumber\\
i_{n+1} &\leftarrow& j_n + 1 \nonumber\\
j_{n+1} &\leftarrow& j_n. \nonumber\\
n &\leftarrow& n+1 \nonumber
\end{eqnarray*}
Go to {\bf Step 2}.

\item {\bf Case 3(b)}: If $\xi_n = h_{j_n}(0)$, then set
\begin{eqnarray*}
y_{j_n}^* &\leftarrow& 0 \nonumber\\
p_{j_n} &\leftarrow& n \nonumber\\
c_{j_n} &\leftarrow& \mathcal{B}^* \nonumber\\
i_{n+1} &\leftarrow& i_n \nonumber\\
j_{n+1} &\leftarrow& j_n-1 \nonumber\\
n &\leftarrow& n+1. \nonumber
\end{eqnarray*}
Go to {\bf Step 2}.

\item {\bf Case 3(c)}: If $\xi_n = \theta_{i_n}^{t}$, for $t = \llbracket  i_n, j_n - 1  \rrbracket $, pick the largest such $t$ and set
\begin{eqnarray*}
y_m^* &\leftarrow& H_m \left(\xi_n \right) \mbox{ for } m \in \llbracket  i_n,t  \rrbracket  \nonumber\\
p_m &\leftarrow& n \mbox{ for } m \in \llbracket  i_n,t  \rrbracket  \nonumber\\
c_m &\leftarrow& \mathcal{C} \mbox{ for } m \in \llbracket  i_n,t-1  \rrbracket  \nonumber\\
c_t &\leftarrow& \mathcal{C}^* \nonumber\\
i_{n+1} &\leftarrow& t + 1 \nonumber\\
j_{n+1} &\leftarrow& j_n. \nonumber\\
n &\leftarrow& n+1. \nonumber
\end{eqnarray*}
Go to {\bf Step 2}. \hspace*{\fill}~\QEDclosed
\end{itemize}
}
\end{algorithm}

{\it Remarks}:
\begin{itemize}
  \item Observe that in each iteration (i.e., a call to Step 3) at least one variable is set. So the algorithm terminates within $L$ steps.
  \item The iterations are indexed by $n$ where $n \in \llbracket  1,N  \rrbracket $. In each iteration, say $n$, the pointers $i_n$ and $j_n$ indicate the start and end variables that are yet to be set. At the end of this iteration, either all the variables are set (Step 3(a)), or the last one alone is set to 0 (Step 3(b)), or the variables with contiguous indices $\llbracket  i_n,t  \rrbracket $ are set (Step 3(c)). The corresponding sets of labels are $\{\mathcal{A}, \mathcal{A}^* \}$, $\{\mathcal{B}^* \}$, and $\{\mathcal{C}, \mathcal{C}^* \}$, respectively.
  \item Suppose $y$ is the vector of production levels of $L$ production units. Let $g_m$ represent the cost of operation for production unit $m \in \llbracket  1,L  \rrbracket$, and $G$ the overall cost. The production levels $y_m$ set in a particular iteration are set to have the same marginal cost $\xi_n$, or are set to operate at capacity. In symbols, $y_m^* = H_m \left(\xi_{p_m} \right) = h_m^{-1} \left( \xi_{p_m} \right) \wedge \beta_m$.
  \item Each iteration requires the evaluation of $\Theta_{i_n}^{j_n}$, and $\theta_{i_n}^{l}$ for $l \in \llbracket  i_n,j_n - 1  \rrbracket $. These are zeros of continuous increasing functions. In Theorem \ref{thm:correctnessOfAlgorithm} below, we provide sufficient conditions under which these zeros exist in each iteration step.
  \item The question of evaluation of these zeros naturally arises. In the specific examples in Section \ref{sec:examples}, we give closed form expressions for $\Theta_{i_n}^{j_n}$ and $\theta_{i_n}^l$. In general, this may not be available and one has to resort to numerical evaluation. However, the observation that the functions are continuous and increasing enables an efficient line search for the zeros. In the proof, we identify $\underline{\theta}$ and $\overline{\theta}$ on either side of the zero (see (\ref{eqn:underlineOverlineTheta}) and (\ref{eqn:underlineOverlineThetaCondition})) that narrows the search window.
\end{itemize}

\vspace*{.25cm}

We now state the main result of the paper.

\vspace*{.25cm}

\begin{thm}
  \label{thm:correctnessOfAlgorithm}
  If $\theta_1^l, l \in \llbracket  1,L-1  \rrbracket $ and $\Theta_1^L$ exist, then the following hold.
  \begin{itemize}
    \item For every iteration $n$ with $i_n \leq j_n$, the quantities $\Theta_{i_n}^{j_n}$ and $\theta_{i_n}^l, l \in \llbracket  i_n, j_n - 1  \rrbracket $ exist.
    \item Algorithm \ref{alg:universal} terminates in $N \leq L$ iterations.
    \item The output of Algorithm \ref{alg:universal} minimizes (\ref{eqn:minimizeSumSeparableConvexFunctions}) under the stated constraints.
  \end{itemize}
\end{thm}

We next state a simple corollary to this result which solves a related problem with additional constraints.

\vspace*{.25cm}

\begin{corollary}
  \label{cor:orderedConditionCorollary}
  If the functions $H_m$ satisfy $H_1 \geq H_2 \geq \cdots \geq H_L$, then under the conditions of Theorem \ref{thm:correctnessOfAlgorithm}, the optimum $y^*$ satisfies $y_1^* \geq y_2^* \geq \cdots \geq y_L^*$.
\end{corollary}

\vspace*{.25cm}

{\it Remark}: We may use Algorithm \ref{alg:universal} to solve the minimization problem with the additional constraints $y_1 \geq y_2 \geq \cdots \geq y_L$ if $H_m$ is point-wise monotone decreasing in the index $m$.

\vspace*{.25cm}

Before we state some properties of the optimum value function, we make some more definitions for convenience.
\begin{itemize}
 \item Observe that if $K > L$, the optimum value defined below depends on the $K$-tuple $\alpha \in \mathbb{R}_+^K$ only through the $L$-tuple $\left( \alpha_1, \cdots, \alpha_{L-1}, \sum_{m=L}^K \alpha_m \right) \in \mathbb{R}_+^L$. For studying the optimum value, we may therefore restrict our attention to $K=L$. Let $\alpha \in \mathbb{R}_+^L$ and define $\mathcal{G}:\mathbb{R}_+^L \rightarrow \mathbb{R} \cup \{+\infty \}$ as follows:
  \[
    \alpha \stackrel{\mathcal{G}}{\mapsto} \mathcal{G}(\alpha) := \inf \left\{ G(y) : y \in \mathbb{R}^L \mbox{ satisfies constraints } (\ref{eqn:positivityBounded})-(\ref{eqn:equalityConstraint}) \right\}
  \]
  We do not place the restrictions (\ref{eqn:commonSense1}) and (\ref{eqn:commonSense2}) on $\alpha$; if the optimization is over an empty set the infimum is taken to be $+\infty$. Clearly $\mathcal{G} > -\infty$ because it is the infimum of a strictly convex function over a bounded convex set, the set being defined by the constraints (\ref{eqn:positivityBounded})-(\ref{eqn:equalityConstraint}).

  \item Define a partial order on $\mathbb{R}_+^K$ as follows. We say $\alpha \succeq \tilde{\alpha}$ if
  \[
     \sum_{m=1}^l \alpha_m \geq \sum_{m=1}^l \tilde{\alpha}_m, ~ l = \llbracket  1,L  \rrbracket ,
  \]
  with equality when $l = L$.

  This partial order is stronger than majorization (see for example Marshall \& Olkin \cite{1979ITMA_MarOlk}) in the sense that if $\alpha \succeq \tilde{\alpha}$, then $\alpha$ majorizes $\tilde{\alpha}$. Loosely speaking, $\alpha \succeq \tilde{\alpha}$ indicates that the components for $\alpha$ are lopsided relative to those of $\tilde{\alpha}$. The proposition below says that lopsidedness increases cost.
\end{itemize}

\vspace*{.25cm}

\begin{proposition} \label{prop:valueFunctionProperties}
The function $\mathcal{G}$ satisfies the following properties:
\begin{itemize}
  \item If $\alpha \succeq \tilde{\alpha}$, then $\mathcal{G}(\alpha) \geq \mathcal{G} \left( \tilde{\alpha} \right)$.
  \item $\mathcal{G}$ is a convex function.
\end{itemize}
\end{proposition}

\vspace*{.25cm}

All the above results are generalizations of those of Viswanath \& Anantharam \cite{200206TIT_VisAna}. The proofs are in Section \ref{sec:proofs}.

\section{Examples}
\label{sec:examples}

We first consider a special case of an example from Bertsekas \cite[Ex. 5.1.2]{2003xxNLP_Ber} that is of interest in optimization of communication systems.

\vspace*{.25cm}

\begin{example}[Vector Gaussian Channel]
\label{example:VectorGaussianChannel}
Consider $L$ channels with noise variances $\sigma_m^2, m \in \llbracket  1,L  \rrbracket $. If power $y_m ~(\geq 0)$ is allocated to channel $m$, the throughput on this channel is $\log \left( 1 + \frac{y_m}{\sigma_m^2} \right)$. Maximize the total throughput
\begin{equation}
  \label{eqn:sumCapacity}
  \sum_{m=1}^L \log \left( 1 + \frac{y_m}{\sigma_m^2} \right)
\end{equation}
subject to a sum power constraint $\sum_{m=1}^L y_m = P$.
\end{example}

\vspace*{.25cm}

The optimal allocation of powers is usually called ``water-filling'' allocation \cite[Sec. 10.4]{1991EIT_CovTho} because it levels the noise-plus-signal power $\sigma_m^2 + y_m^*$ in the channels subject to the sum power constraint, possibly leaving out a few of the noisiest dimensions. We now arrive at this solution using our algorithm.

Without loss of generality, we arrange the indices so that
\begin{equation}
  \label{eqn:orderingWaterFilling}
  \sigma_1^2 \leq \sigma_2^2 \leq \cdots \leq \sigma_L^2.
\end{equation}
Set $g_m(x) = -\log \left( 1+\frac{x}{\sigma_m^2} \right)$, $(a_m, b_m) = \left(-\sigma_m^2, +\infty\right)$, $\beta_m = +\infty$, $K=L$, $\alpha_m = 0$ for $m \in \llbracket  1,L-1  \rrbracket $, and $\alpha_L = P$. It is easy to verify that $g_m, m \in \llbracket  1,L  \rrbracket $ satisfy all the conditions laid out in Section \ref{sec:problemDescription} and that
\begin{equation}
  \label{eqn:waterFillingHm}
  H_m(\theta) = - \theta^{-1} - \sigma_m^2
\end{equation}
with domain $\mathbb{E}_m = (-\infty, 0)$. Consequently, $\theta_1^l$, the solution to (\ref{eqn:smallTheta}), is given by
\begin{equation}
  \label{eqn:explicitSmallTheta}
  \theta_1^l = \frac{-l}{\sum_{m=1}^l \sigma_m^2}, l \in \llbracket  1, L-1  \rrbracket ,
\end{equation}
and
\begin{equation}
  \label{eqn:explicitBigTheta}
  \Theta_1^L = \frac{-L}{\sum_{m=1}^L \sigma_m^2 + P}.
\end{equation}
Theorem \ref{thm:correctnessOfAlgorithm} therefore indicates that Algorithm \ref{alg:universal} is applicable. $\theta_1^l$ and $\Theta_1^j$ are similarly identified for $1 \leq l < j \leq L$. The ordering in (\ref{eqn:orderingWaterFilling}) implies that $\Theta_1^j \geq \theta_1^l$ for all $l \in \llbracket  1,j-1  \rrbracket $. An execution of Algorithm \ref{alg:universal} therefore results in the following: identify the largest $j$ such that $\Theta_1^j \geq h_j(0) = -\sigma_j^{-2}$, or equivalently, $\sigma_j^2 \leq -1/\Theta_1^j$, set $y_m^* = 0$ for $m \in \llbracket  j+1, L  \rrbracket $, and set $y_m = H_m \left(\Theta_1^j \right)$ for $m \in \llbracket  1,j  \rrbracket $. From (\ref{eqn:waterFillingHm}), we see that $y_m^* + \sigma_m^2 = -1/\Theta_1^j$, the water level, for $m \in \llbracket  1,j  \rrbracket $. \hspace*{\fill}~\QEDclosed

\vspace*{.25cm}

Our second example is closely related to Example \ref{example:VectorGaussianChannel} and is from Viswanath \& Anantharam \cite{200206TIT_VisAna}. It evaluates the sum throughput in a multi-user setting. Omitting the details of the reduction, we present only the mathematical abstraction.

\vspace*{.25cm}

\begin{example}[Sum Capacity]
\label{example:sumCapacity}
Let (\ref{eqn:orderingWaterFilling}) hold. Let $g_m(x) = - \log \left( 1 + \frac{x}{\sigma_m^2} \right), m \in \llbracket  1,L  \rrbracket ,$ as in Example \ref{example:VectorGaussianChannel}, and let $\alpha_1 \geq \alpha_2 \geq \cdots \geq \alpha_K \geq 0$. Minimize (\ref{eqn:minimizeSumSeparableConvexFunctions}) subject to the constraints in Section \ref{sec:problemDescription}.
\end{example}

\vspace*{.25cm}

The difference between Examples \ref{example:VectorGaussianChannel} and \ref{example:sumCapacity} is that the ascending constraints now apply. Existence of $\theta_1^l, l \in \llbracket 1,L-1\rrbracket$ and $\Theta_1^L$ follows as in the previous example (see (\ref{eqn:example2Existence}) below), and Algorithm \ref{alg:universal} can be used to solve the minimization. Analogous to (\ref{eqn:explicitSmallTheta}) and (\ref{eqn:explicitBigTheta}), $\theta_i^l$ and $\Theta_i^j$ are given by
\begin{equation}
  \label{eqn:example2Existence}
  \theta_i^l = \frac{-(l-i+1)}{\sum_{m=i}^l \sigma_m^2} \mbox{ and }
  \Theta_i^j = \frac{-(j-i+1)}{\sum_{m=i}^j \sigma_m^2 + P}.
\end{equation}
for $1 \leq i \leq l < j \leq L$. At iteration step $n$, given $i_n$ and $j_n$, the assignment for $\xi_n$ is explicitly given by
\begin{eqnarray}
  \lefteqn{  \xi_n = \max \left\{ \frac{-(j_n - i_n + 1)}{\sum_{m = i_n}^{j_n} \sigma_m^2 + \sum_{m = i_n}^{K} \alpha_m}, \frac{-1}{\sigma_{j_n}^2}, \right. } \nonumber \\
  && ~~~~~~~~~~~~~~~~~~~~~~\left. \frac{-(l - i_n + 1)}{\sum_{m = i_n}^l \sigma_m^2 + \sum_{m = i_n}^{l} \alpha_m}; l \in \llbracket  i_n,j_n - 1  \rrbracket   \right\}.
  \label{eqn:VisAnaSumCapacityXin}
\end{eqnarray}
Viswanath \& Anantharam \cite{200206TIT_VisAna} give the same condition in terms of $-1/\xi_n$. The optimal allocation at this step sets either $y_{j_n}^* = 0$ indicating a rejection of the noisy channel index $j_n$, or $y_m^* = H_m \left(\xi_n \right)$, i.e., $y_m^* + \sigma_m^2 = -\xi_n^{-1}$ for a contiguous set of channels starting from index $i_n$. This optimal allocation thus partitions the channels into sets of contiguous channels, with each partition having its own water level.

\vspace*{.25cm}

{\it Remark}: Viswanath \& Anantharam make an incorrect claim in \cite[Appendix A.5]{200206TIT_VisAna} that the specific algorithm with $\xi_n$ set via (\ref{eqn:VisAnaSumCapacityXin}) puts out the optimal $y^*$ whenever $g_m, m \in \llbracket  1,L  \rrbracket $ are of the form $g_m(x) = f \left(1 + \frac{x}{\sigma_m^2}\right)$ and $f : \mathbb{R}_+ \rightarrow \mathbb{R}$ is a continuous, increasing, strictly concave function. Their proof works only for some special cases. In particular, it works for $f(x) = \log x$ as in Example \ref{example:sumCapacity} above. The error in their proof can be traced to an incomplete proof for the case when $L=2$ in \cite[p.1309]{200206TIT_VisAna}; the validity of their statement for $L=2$ holds only in some special cases, $f(x) = \log(x)$ being one of them. Of course, Algorithm \ref{alg:universal} with the correct $\xi_n$ based on the functions $g_m, m \in \llbracket  1,L  \rrbracket $ will yield the optimal $y^*$. \hspace*{\fill}~\QEDclosed

\vspace*{.25cm}

Our third example is also taken from Viswanath \& Anantharam \cite{200206TIT_VisAna} and evaluates the minimum power required to meet a quality-of-service constraint. It serves to illustrate the use of Corollary \ref{cor:orderedConditionCorollary}. The mathematical abstraction is as follows.

\vspace*{.25cm}

\begin{example}
  \label{example:minPower}
  Let (\ref{eqn:orderingWaterFilling}) hold. Let $g_m:(-\infty,1)\rightarrow \mathbb{R}_+$ be defined as $g_m (x) = \frac{\sigma_m^2}{1-x}, m \in \llbracket  1,L  \rrbracket $. Let   $\alpha_1 \geq \alpha_2 \geq \cdots \geq \alpha_K$. Minimize (\ref{eqn:minimizeSumSeparableConvexFunctions}) subject to the constraints in Section \ref{sec:problemDescription} and the additional set of constraints $y_1 \geq y_2 \geq \cdots \geq y_L$.
\end{example}

\vspace*{.25cm}

Let us first solve the problem ignoring the constraint $y_1 \geq y_2 \geq
\cdots \geq y_L$. Observe that $(a_m,b_m) = (-\infty,1)$ for all $m$ and all the conditions outlined in Section \ref{sec:problemDescription} are satisfied by the given set of functions. Furthermore, it is easy to verify that $\mathbb{E}_m = (0,\infty)$ and $H_m(\theta) = 1 - \frac{\sigma_m}{\sqrt{\theta}}$. The quantities $\theta_1^l$ and $\Theta_1^L$ exist if $\sum_{m=1}^l \alpha_m < l$ and $\sum_{m=1}^K \alpha_m< L$. More generally, for $1 \leq i \leq l < j \leq L$, $\theta_i^l$ and
$\Theta_i^j$ are given by
\[
  \theta_i^l = \left( \frac{\sum_{m=i}^l \sigma_m}{l - i + 1 - \sum_{m=i}^l \alpha_m} \right)^2
\]
and
\[
  \Theta_i^j = \left( \frac{\sum_{m=i}^j \sigma_m}{j - i + 1 - \sum_{m=i}^K \alpha_m} \right)^2.
\]
Theorem \ref{thm:correctnessOfAlgorithm} then assures us that Algorithm \ref{alg:universal} yields the optimum solution. At iteration step $n$, given $i_n$ and $j_n$, the assignment for $\xi_n$ can once again be made more explicit and our algorithm reduces to the second algorithm of Viswanath \& Anantharam. Observe now that (\ref{eqn:orderingWaterFilling}) implies that $H_1 \geq H_2 \geq \cdots \geq H_L$ so that the optimal $y^*$ put out by Algorithm \ref{alg:universal} also satisfies $y_1^* \geq y_2^* \geq \cdots \geq y_L^*$. \hspace*{\fill}~\QEDclosed

\vspace*{.25cm}

Our final example illustrates the handling of the upper bound constraint. This is another special case of the example from Bertsekas \cite[Ex. 5.1.2]{2003xxNLP_Ber} that arises in a power optimization problem for sensor networks (see Zacharias \& Sundaresan \cite{200707TITarXiv_ZacSun}).

\vspace*{.25cm}

\begin{example}
  \label{example:minVariance}
  Let $g_m(x) = \frac{1}{2} x^2, m \in \llbracket  1,L  \rrbracket $, $K = L$, $\alpha_m = 0, m \in \llbracket  1,L-1  \rrbracket $, $\alpha_L = \alpha > 0$, and $\beta_m \in (0, \infty)$ for $m \in \llbracket  1,L  \rrbracket $. Further, order the indices so that $\beta_1 \leq \beta_2 \leq \cdots \leq \beta_L$. Minimize (\ref{eqn:minimizeSumSeparableConvexFunctions}) under this setting.
\end{example}

\vspace*{.25cm}

Once again, all conditions outlined in Section \ref{sec:problemDescription} hold. It is easy to verify that $H_m(\theta) = \theta \wedge \beta_m$. The function $\sum_{m=1}^l H_m$ is a piece-wise linear continuous function passing through the origin with slope $l$ in $(-\infty, \beta_1)$, slope $(l - 1)$ in $(\beta_1, \beta_2)$, and so on, and zero-slope in $(\beta_l, +\infty)$. Clearly $\theta_1^l = 0, l \in \llbracket  1,L-1  \rrbracket $. We assume that $\Theta_1^L$ exists which is equivalent to $\alpha \leq \sum_{m=1}^L \beta_L$. Yet again, Theorem \ref{thm:correctnessOfAlgorithm} assures us that Algorithm \ref{alg:universal} is applicable.

An application of Algorithm \ref{alg:universal} results in the following. Identify the unique $k$ such that $\alpha \in [a_k, a_{k+1})$ where
\[
  a_k := (L-k) \beta_k + \sum_{m=1}^k \beta_m.
\]
(It is easy to see that $a_k \leq a_{k+1}$). Then
\[
  \Theta_1^L = \frac{ \alpha - \sum_{m=1}^k \beta_m }{L-k}.
\]
Moreover, $y_m^* = H_m \left( \Theta_1^L \right) = \beta_m$ for $m \in \llbracket  1,k  \rrbracket $. For $m = \llbracket  k+1, L  \rrbracket $, the values are suitably lowered from their upper bounds. Note that this assignment is completed in just one iteration of Algorithm \ref{alg:universal}. \hspace*{\fill}~\QEDclosed

\section{Proofs}
\label{sec:proofs}

\subsection{Preliminaries}
\label{subsec:preliminaries}

We first prove some facts on the individual cases.

\vspace*{.25cm}

\begin{lemma} \label{lem:3b}
Suppose that in iteration $n$, the quantities $\theta_{i_n}^l, l \in \llbracket  i_n, j_n - 1  \rrbracket $ and $\Theta_{i_n}^{j_n}$ exist. Suppose further that $\xi_n = h_{j_n} (0)$ and Step 3(b) is executed. Then the following hold.
\begin{itemize}
  \item The quantities $\theta_{i_{n+1}}^l, l \in \llbracket  i_{n+1}, j_{n+1}-1  \rrbracket $ and $\Theta_{i_{n+1}}^{j_{n+1}}$ exist.
  \item $\xi_n \geq \xi_{n+1}$.
\end{itemize}
\end{lemma}

\vspace*{.25cm}

\begin{proof}
Given that $\xi_n = h_{j_n} (0)$ and Step 3(b) is executed, we see that $i_{n+1} = i_n$ and $j_{n+1} = j_n - 1$. We may assume $i_{n+1} \leq j_{n+1}$; otherwise there is nothing to prove. Since the start pointer does not change and the end pointer decrements by 1, it is clear that $\theta_{i_{n+1}}^l = \theta_{i_n}^l$ for $l=\llbracket  i_{n+1},j_{n+1}-1  \rrbracket $ because $\theta_{i_{n+1}}^l$ and $\theta_{i_n}^l$ are zeros of identical functions for the indicated values of $l$.

The question of existence now reduces to that of only $\Theta_{i_{n+1}}^{j_{n+1}}$ (recall definition in (\ref{eqn:bigTheta})). First observe that if Step 3(b) is executed, we must have $\xi_n = h_{j_n}(0) > \Theta_{i_n}^{j_n}$, and therefore
\begin{equation}
  \label{eqn:3bOverlineTheta}
  0 = h_{j_n}^{-1} \left( h_{j_n}(0) \right) = h_{j_n}^{-1} \left( \xi_n \right) \geq H_{j_n}\left( \xi_n \right) \geq H_{j_n} \left( \Theta_{i_n}^{j_n} \right),
\end{equation}
where the last inequality follows because $H_{j_n}$ is increasing. Consequently, we must have
\begin{eqnarray*}
  \sum_{m=i_{n+1}}^K \alpha_m & = & \sum_{m=i_n}^K \alpha_m
    = \sum_{m={i_n}}^{j_n} H_m \left( \Theta_{i_n}^{j_n} \right)  ~~ \left( \mbox{from definition of } \Theta_{i_n}^{j_n} \right) \\
    & = & \sum_{m={i_n}}^{j_n-1} H_m \left( \Theta_{i_n}^{j_n} \right) + H_{j_n} \left( \Theta_{i_n}^{j_n} \right) \\
    & \leq & \sum_{m={i_n}}^{j_n-1} H_m \left( \Theta_{i_n}^{j_n} \right) + 0 ~~~~~~~~~~~~ \mbox{(from (\ref{eqn:3bOverlineTheta}))} \\
    & = & \sum_{m={i_{n+1}}}^{j_{n+1}} H_m \left( \Theta_{i_n}^{j_n} \right).
\end{eqnarray*}
So we may take $\overline{\theta} = \Theta_{i_n}^{j_n}$ in (\ref{eqn:underlineOverlineTheta}). On the lower side, we may simply use $\underline{\theta} = h_1(0)$. However, we can find a tighter bound from the following sequence of inequalitites
\begin{eqnarray*}
  \sum_{m=i_{n+1}}^{j_{n+1}} H_m \left( \theta_{i_n}^{j_n-1} \right) & = & \sum_{m=i_n}^{j_n-1} H_m \left( \theta_{i_n}^{j_n-1} \right)
   = \sum_{m=i_n}^{j_n-1} \alpha_m ~~ \left( \mbox{from definition of } \theta_{i_n}^{j_n-1} \right) \\
  & < &\sum_{m=i_n}^K \alpha_m ~~ \left( \mbox{ from (\ref{eqn:commonSense2})}\right) \\
  & = & \sum_{m=i_{n+1}}^K \alpha_m;
\end{eqnarray*}
i.e., we may take $\underline{\theta} = \theta_{i_n}^{j_n-1}$ in (\ref{eqn:underlineOverlineTheta}). $\Theta_{i_{n+1}}^{j_{n+1}}$ therefore exists and
\begin{equation}
  \label{eqn:3bXin}
  \theta_{i_n}^{j_n-1} < \Theta_{i_{n+1}}^{j_{n+1}} \leq \Theta_{i_n}^{j_n}.
\end{equation}
This establishes the existence part of the Lemma.

To establish $\xi_n \geq \xi_{n+1}$, we simply observe that $\xi_n$ is at least as large as all the candidates that determine $\xi_{n+1}$. Indeed, $\xi_{n} = h_{j_n}(0) \geq h_{j_n-1}(0) = h_{j_{n+1}}(0)$. Next $\xi_n \geq \theta_{i_n}^l = \theta_{i_{n+1}}^l$ for $l \in \llbracket  i_{n+1},j_{n+1}-1  \rrbracket $, and finally, $\xi_n > \Theta_{i_n}^{j_n} \geq \Theta_{i_{n+1}}^{j_{n+1}}$ by the right side inequality of (\ref{eqn:3bXin}). This completes the proof of the lemma.
\end{proof}

\vspace*{.25cm}

\begin{lemma} \label{lem:3c}
Suppose that in iteration $n$, the quantities $\theta_{i_n}^l, l \in \llbracket  i_n,j_n - 1  \rrbracket $ and $\Theta_{i_n}^{j_n}$ exist. Suppose further that $\xi_n = \theta_{i_n}^t$ for some $t \in \llbracket  i_n, j_n - 1  \rrbracket $, and Step 3(c) is executed. Then the following hold.
\begin{itemize}
  \item $y_m^* \in [0, \beta_m], m \in \llbracket  i_n,t  \rrbracket $.
  \item $\sum_{m=i_n}^l y_m^* \geq \sum_{m=i_n}^l \alpha_m, l \in \llbracket  i_n,t  \rrbracket $, with equality when $l = t$.
  \item The quantities $\theta_{i_{n+1}}^l, l \in \llbracket  i_{n+1},j_{n+1}-1  \rrbracket $ and $\Theta_{i_{n+1}}^{j_{n+1}}$ exist.
  \item $\xi_n \geq \xi_{n+1}$.
\end{itemize}
\end{lemma}

\vspace*{.25cm}

\begin{proof}
Note that in this case $t$ is chosen to be the largest one in $\llbracket  i_n, j_n-1  \rrbracket $ that satisfies $\xi_n = \theta_{i_n}^t$. Step 3(c) is executed; therefore $\theta_{i_n}^t \geq h_{j_n}(0) \geq h_m(0)$ for $m = \llbracket  i_n, t  \rrbracket $. The assignment for $y_m^*$ in the algorithm satisfies
\[
  y_m^* = H_m \left( \theta_{i_n}^t \right) \geq H_m \left( h_m(0) \right) = h_m^{-1} \left( h_m(0) \right) \wedge \beta_m = 0 \wedge \beta_m = 0.
\]
That $y_m^* \leq \beta_m$ is obvious from the definition of $H_m$. This proves the upper and lower bound constraints on $y_m^*$.

To show that the ascending constraints (with the sum starting from $i_n$) hold for $l = \llbracket  i_n, t  \rrbracket $, observe that $\theta_{i_n}^t \geq \theta_{i_n}^l$ and the increasing property of $H_m$ imply
\[
  \sum_{m=i_n}^l y_m^* = \sum_{m=i_n}^l H_m \left( \theta_{i_n}^t \right) \geq \sum_{m=i_n}^l H_m \left( \theta_{i_n}^l \right) = \sum_{m=i_n}^l \alpha_m,
\]
with equality when $l = t$.

We next consider existence of $\theta_{i_{n+1}}^l$. Recall that in Step 3(c), $i_{n+1} = t+1$ and $j_{n+1} = j_n$. Fix $l \in \llbracket  i_{n+1}, j_{n+1}-1  \rrbracket $. We simply set $\underline{\theta} = h_1(0)$ in (\ref{eqn:underlineOverlineTheta}). Moreover,
\begin{eqnarray*}
  \sum_{m=i_{n+1}}^{l} \alpha_m & = & \sum_{m=i_n}^{l} \alpha_m - \sum_{m=i_n}^{t} \alpha_m \\
  & = & \sum_{m=i_n}^{l} H_m \left( \theta_{i_n}^l \right) - \sum_{m=i_n}^{t} H_m \left( \theta_{i_n}^t \right) ~\left( \mbox{from the definitions of } \theta_{i_n}^l \mbox{ and } \theta_{i_n}^t \right)\\
  & \leq & \sum_{m=i_n}^{l} H_m \left( \theta_{i_n}^l \right) - \sum_{m=i_n}^{t} H_m \left( \theta_{i_n}^l \right) ~(\mbox{because } H_m \mbox{ is increasing})\\
  & = & \sum_{m=i_{n+1}}^{l} H_m \left( \theta_{i_n}^l \right),
\end{eqnarray*}
and therefore we may set $\overline{\theta} = \theta_{i_n}^l$ in (\ref{eqn:underlineOverlineTheta}). $\theta_{i_{n+1}}^l$ therefore exists and
\begin{equation}
  \label{eqn:3cOverlineSmallTheta}
  h_1(0) \leq \theta_{i_{n+1}}^l \leq \theta_{i_n}^l.
\end{equation}
The same argument (mutatis mutandis to account for the sum of $\alpha_m$ up to $K$) establishes the existence of $\Theta_{i_{n+1}}^{j_{n+1}}$ and that
\begin{equation}
  \label{eqn:3cOverlineBigTheta}
  h_1(0) \leq \Theta_{i_{n+1}}^{j_{n+1}} \leq \Theta_{i_n}^{j_n}.
\end{equation}
Finally, to show $\xi_n \geq \xi_{n+1}$, observe that $\xi_n \geq h_{j_n}(0) = h_{j_{n+1}}(0)$, $\xi_n \geq \Theta_{i_n}^{j_n} \geq \Theta_{i_{n+1}}^{j_{n+1}}$, and $\xi_n \geq \theta_{i_n}^l \geq \theta_{i_{n+1}}^l, l \in \llbracket  i_{n+1}, j_{n+1}-1  \rrbracket $. The last two facts follow from (\ref{eqn:3cOverlineBigTheta}) and (\ref{eqn:3cOverlineSmallTheta}). So $\xi_n$ is at least as large as all the candidates that determine $\xi_{n+1}$, i.e., $\xi_n \geq \xi_{n+1}$, and the proof is complete.
\end{proof}

\vspace*{.25cm}

\begin{lemma} \label{lem:3a}
Suppose that in iteration $n$, the quantities $\theta_{i_n}^l, l \in \llbracket  i_n,j_n - 1  \rrbracket $ and $\Theta_{i_n}^{j_n}$ exist. Suppose further that $\xi_n = \Theta_{i_n}^{j_n}$. Then the following hold.
\begin{itemize}
  \item $y_m^* \in [0, \beta_m], m \in \llbracket  i_n,j_n  \rrbracket $.
  \item $\sum_{m=i_n}^l y_m^* \geq \sum_{m=i_n}^l \alpha_m, l = \llbracket  i_n,j_n  \rrbracket $, with equality when $l = j_n$.
\end{itemize}
\end{lemma}

\vspace*{.25cm}

\begin{proof}
Under the hypotheses, Step 3(a) is executed. The proofs of the statements are identical to the proofs of the first two parts of Lemma \ref{lem:3c} and is omitted.
\end{proof}

\vspace*{.25cm}

\begin{proposition}
  \label{prop:correctness}
  If $\theta_1^l, l \in \llbracket  1, L-1  \rrbracket $ and $\Theta_1^L$ exist, the following statements hold.
  \begin{itemize}
     \item For every iteration step $n$ with $i_n \leq j_n$, the quantities $\Theta_{i_n}^{j_n}$ and $\theta_{i_n}^l, l \in \llbracket  i_n, j_n - 1  \rrbracket $ exist.
     \item Algorithm \ref{alg:universal} terminates in $N \leq L$ steps.
     \item The output $y^*$ of Algorithm \ref{alg:universal} is feasible.
     \item $\xi_1 \geq \xi_2 \geq \cdots \geq \xi_N$.
     \item In iteration $N$, Step 3(a) is executed.
  \end{itemize}
\end{proposition}

\vspace*{.25cm}

\begin{proof}
The key issue is the existence of $\theta_{i_n}^l$ and $\Theta_{i_n}^{j_n}$ in Step 3 of each iteration. The hypothesis of this Proposition resolves the issue for $n=1$. Lemmas \ref{lem:3b}, \ref{lem:3c}, and \ref{lem:3a} resolve the issue for subsequent iterations via induction. The first statement follows.

At least one variable is set in every iteration. The algorithm thus runs to completion in $N \leq L$ iterations, and the second statement holds.

The third and fourth statements also follow from Lemmas \ref{lem:3b}, \ref{lem:3c}, \ref{lem:3a}, and induction.

We now argue that Step 3(a) is executed in the last iteration. If this is not the case, the last iteration must be Step 3(b). This implies $i_N = j_N$ and $h_{j_N}(0) > \Theta_{j_N}^{j_N}$. The latter inequality and the definition of $\Theta_{j_N}^{j_N}$ yield
\[
  \sum_{m=j_N}^K \alpha_m = H_{j_N} \left( \Theta_{j_N}^{j_N} \right) \leq h_{j_N}^{-1} \left( \Theta_{j_N}^{j_N} \right) \leq h_{j_N}^{-1} \left( h_{j_N}(0) \right) = 0
\]
contradicting our assumption (\ref{eqn:commonSense2}) that $\sum_{m=L}^K \alpha_m>0$.
\end{proof}

\subsection{Proof of Theorem \ref{thm:correctnessOfAlgorithm}}
Proposition \ref{prop:correctness} implies the first two statements of Theorem \ref{thm:correctnessOfAlgorithm}. We now proceed to show the optimality of $y^*$ to complete the proof of Theorem \ref{thm:correctnessOfAlgorithm}.

We use the Karush-Kuhn-Tucker (KKT) conditions (see for example \cite[Sec. 3.3]{2003xxNLP_Ber}) to show that the vector put out by the algorithm is a stationary point of a Lagrangian function with appropriately chosen Lagrange multipliers. The Lagrangian function for the problem is
\begin{eqnarray}
  \lefteqn{ \sum_{m=1}^L g_m(y_m) + \sum_{m=1}^L \lambda_m^{(1)} \left( - y_m \right) + \sum_{m=1}^L \lambda_m^{(2)} \left( y_m - \beta_m \right) }
  \nonumber \\
  \label{eqn:lagrangian}
  & & ~~~~+ \sum_{l=1}^{L-1} \lambda_l^{(3)} \left( -\sum_{m=1}^l y_m + \sum_{m=1}^l \alpha_m \right) + \mu \left( -\sum_{m=1}^L y_m + \sum_{m=1}^K \alpha_m \right)
\end{eqnarray}
where $\lambda_m^{(1)}$ is the Lagrange multiplier that relaxes the positivity constraint $-y_m \leq 0$, $\lambda_m^{(2)}$ relaxes the upper bound constraint $y_m - \beta_m \leq 0$, $\lambda_m^{(3)}$ the ascending constraint (\ref{eqn:ladderConstraint}), and $\mu$ the equality constraint (\ref{eqn:equalityConstraint}).
The KKT necessary and sufficient conditions for optimality of this convex optimization problem are given by:
\begin{eqnarray}
  \label{eqn:kkt-lb}
  \lambda_m^{(1)} y_m & = 0, & m = \llbracket  1,L  \rrbracket  \\
  \label{eqn:kkt-ub}
  \lambda_m^{(2)} \left( y_m - \beta_m \right) & = 0, & m = \llbracket  1,L  \rrbracket  \\
  \label{eqn:kkt-ladderIneq}
  \lambda_l^{(3)} \left( \sum_{m=1}^l y_m - \sum_{m=1}^l \alpha_m \right) & = 0, & l = \llbracket  1,L-1  \rrbracket  \\
  \label{eqn:kkt-nonnegativity}
  \lambda_m^{(1)} \geq 0, \lambda_m^{(2)} \geq 0, ~~m = \llbracket  1,L  \rrbracket , \mbox{ and } \lambda_l^{(3)} & \geq 0, & l = \llbracket  1,L-1  \rrbracket , \\
  \label{eqn:kkt-derivative}
  h_m(y_m) - \lambda_m^{(1)} + \lambda_m^{(2)} - \sum_{l=m}^{L-1} \lambda_m^{(3)} - \mu & = 0, & m = \llbracket  1, L  \rrbracket .
\end{eqnarray}
Conditions (\ref{eqn:kkt-lb}), (\ref{eqn:kkt-ub}), and (\ref{eqn:kkt-ladderIneq}) are the complementary slackness conditions, (\ref{eqn:kkt-nonnegativity}) are the positivity conditions, and (\ref{eqn:kkt-derivative}) identifies a stationary point for the Lagrangian function. We now choose appropriate values for the Lagrange multipliers and verify the KKT conditions.

First, let
\begin{equation}
  \label{eqn:lambda1}
  \lambda_m^{(1)} = \left\{
    \begin{array}{ll} \xi_{p_m} - \xi_N, & \mbox{ if } c_m = \mathcal{B}^*, \\
                      0, & \mbox{ otherwise}.
    \end{array} \right.
\end{equation}
Recall that $p_m$ is the iteration number in which variable $y_m$ was set, and that $c_m = \mathcal{B}^*$ whenever Step 3(b) is executed, i.e., $y_m^* = 0$. From the assignment in (\ref{eqn:lambda1}), $\lambda_m^{(1)} \neq 0$ implies that $c_m = \mathcal{B}^*$ and therefore $y_m^* = 0$. Thus the complementary slackness condition (\ref{eqn:kkt-lb}) is satisfied for $m = \llbracket  1,L  \rrbracket $.

Second, let
\begin{equation}
  \label{eqn:lambda2}
  \lambda_m^{(2)} = \left\{
    \begin{array}{ll} 0, & \mbox{ if } c_m = \mathcal{B}^*, \\
                      \xi_{p_m} - h_m \left( H_m \left( \xi_{p_m} \right) \right), & \mbox{ otherwise}.
    \end{array} \right.
\end{equation}
If $\lambda_m^{(2)} \neq 0$, then from (\ref{eqn:lambda2}) we have $\xi_{p_m} \neq h_m \left( H_m \left( \xi_{p_m} \right) \right)$. From the strictly increasing property of $h_m$ and the definition of $H_m$, we have
\begin{equation}
  \label{eqn:kkt-ub-consequences}
  h_m \left( H_m \left( \xi_{p_m} \right) \right) = h_m \left( h_m^{-1} \left( \xi_{p_m} \right) \wedge \beta_m \right) = h_m \left( h_m^{-1} \left( \xi_{p_m} \right) \right) \wedge h_m \left( \beta_m \right) = \xi_{p_m} \wedge h_m \left( \beta_m \right),
\end{equation}
so that $h_m \left(H_m \left(\xi_{p_m} \right) \right) \neq \xi_{p_m}$ implies that $H_m \left( \xi_{p_m} \right)$ must have saturated to $\beta_m$, i.e., $y_m^* = H_m \left( \xi_{p_m} \right) = \beta_m$. The complementary slackness conditions (\ref{eqn:kkt-ub}) are therefore fulfilled.

Third, for $l=\llbracket  1,L-1  \rrbracket $ let
\begin{equation}
  \label{eqn:lambda3}
  \lambda_l^{(3)} = \left\{
    \begin{array}{ll} \xi_{p_l} - \xi_{p_k}, & \mbox{ if } c_l = \mathcal{C}^*, \\
                      0, & \mbox{ otherwise},
    \end{array} \right.
\end{equation}
where
\begin{equation}
 \label{eqn:lambda3pkdefn}
p_k := \min \left\{ p_m : m \in \llbracket 1,L \rrbracket, p_m > p_l, c_m \in \left\{ \mathcal{C}^{*},\mathcal{A}^{*} \right\} \right\}.
\end{equation}

The last iteration is always via Step 3(a) (Proposition \ref{prop:correctness}). Thus, when $c_l = \mathcal{C}^*$, there is a later iteration that executes Step 3(a) which implies that the set in (\ref{eqn:lambda3pkdefn}) is nonempty and that the assignment (\ref{eqn:lambda3}) is well-defined. Suppose $\lambda_l^{(3)} \neq 0$. Then $c_l = \mathcal{C}^*$, an asterisked assignment. The second statement of Lemma \ref{lem:3c} therefore ensures that the ascending constraint is satisfied with equality for this $l$. The complementary slackness condition (\ref{eqn:kkt-ladderIneq}) is thus fulfilled for $l = \llbracket  1,L-1  \rrbracket $.

The assignment of $\lambda_l^{(3)}$ in (\ref{eqn:lambda3}) can be equivalently expressed as
\begin{equation}
  \label{eqn:lambda3equiv}
  \lambda_l^{(3)} = \left\{
    \begin{array}{ll} \sum_{n=p_l}^{p_k-1}\xi_{n} - \xi_{n+1}, & \mbox{ if } c_l = \mathcal{C}^*, \\
                      0, & \mbox{ otherwise},
    \end{array} \right.
\end{equation}
where $p_k$ is given by (\ref{eqn:lambda3pkdefn}). This will be useful in verifying (\ref{eqn:kkt-derivative}).

Finally, we set $\mu = \xi_N$.

The Lagrange multiplier assignments in (\ref{eqn:lambda1}), (\ref{eqn:lambda2}), and (\ref{eqn:lambda3}) are positive. Indeed, the positivity in (\ref{eqn:lambda1}) and (\ref{eqn:lambda3}) follow from the monotonicity property $\xi_n \geq \xi_{n+1}, n=\llbracket  1,N-1  \rrbracket $ (Proposition \ref{prop:correctness}). The positivity of $\lambda_m^{(2)}$ follows from
\[
  h_m \left( H_m \left(\xi_{p_m} \right) \right) \leq h_m \left( h_m^{-1} \left(\xi_{p_m} \right) \right) = \xi_{p_m}.
\]

All that remains is to verify (\ref{eqn:kkt-derivative}). To do this, first consider $m > j_N$. Then the assignments $y_m^*=0$ and $y_l^*=0, l=\llbracket  m+1,L-1  \rrbracket $ are via Step 3(b); therefore $\xi_{p_l} = h_l(0)$ and $c_l = \mathcal{B}^*$. The latter implies $\lambda_m^{(1)} = \xi_{p_m} - \xi_{N}$, $\lambda_m^{(2)} = 0$, and $\lambda_l^{(3)} = 0$ for $l = \llbracket  m,L-1  \rrbracket $. Substitution of these assignments in (\ref{eqn:kkt-derivative}) yields
\[
  h_m(0) - \lambda_m^{(1)} + \lambda_m^{(2)} - \sum_{l=m}^{L-1} \lambda_l^{(3)} - \mu = \xi_{p_m} - \left(\xi_{p_m} - \xi_N \right) + 0 - 0 - \xi_{N} = 0.
\]

Now consider $m \in \llbracket  1, j_N  \rrbracket $ and $p_m < N$, i.e., variable $y_m$ is not set in the last iteration. Then $c_m \in \{ \mathcal{C}, \mathcal{C}^* \}$, and thus $c_m \neq \mathcal{B}^*$. Substitution of (\ref{eqn:lambda1}), (\ref{eqn:lambda2}), and (\ref{eqn:lambda3}) in (\ref{eqn:kkt-derivative}) yields
\begin{eqnarray}
  \nonumber
  \lefteqn{ h_m \left( y_m^* \right) - \lambda_m^{(1)} + \lambda_m^{(2)} - \sum_{l=m}^{L-1} \lambda_l^{(3)} - \mu } \\
  \nonumber
  & = & h_m \left( H_m \left(\xi_{p_m} \right) \right) - 0 + \left( \xi_{p_m} - h_m \left( H_m \left( \xi_{p_m} \right) \right) \right) - \sum_{l=m}^{L-1} \lambda_l^{(3)} - \xi_{N} \\
  \label{eqn:Lambda3Sum}
  & = & \xi_{p_m} - \xi_N - \sum_{l=m}^{L-1} \lambda_l^{(3)}\\
  \label{eqn:splitLambda3Sum}
  & = & \xi_{p_m} - \xi_N - \sum_{n=p_m}^{N-1} \left( \xi_n - \xi_{n+1} \right) \\
  \nonumber
  & = & \xi_{p_m} - \xi_N - \left( \xi_{p_m} - \xi_{N} \right) \\
  \nonumber
  & = & 0.
\end{eqnarray}
In the above sequence of inequalities, (\ref{eqn:splitLambda3Sum}) holds because of the following. In (\ref{eqn:Lambda3Sum}), the summation over $l$ has only one nonzero entry per iteration, i.e., whenever $c_l = \mathcal{C}^*$. We may therefore sum over the iteration index $n$ instead of the variable index $l$. Iterations $p_m$ to $N-1$ involve the execution of either Step 3(b) or Step 3(c).  Substitution of (\ref{eqn:lambda3equiv}) in (\ref{eqn:Lambda3Sum}) then results in (\ref{eqn:splitLambda3Sum}).

Lastly, consider $m \in \llbracket  1, j_N  \rrbracket $ and $p_m = N$, i.e., $y_m$ is assigned in the last iteration. From Proposition \ref{prop:correctness}, Step 3(a) is executed in this iteration, and therefore $c_m \in \{ \mathcal{A}, \mathcal{A}^* \}$. Then
\begin{eqnarray*}
  \lefteqn{ h_m \left(y_m \right) - \lambda_m^{(1)} + \lambda_m^{(2)} - \sum_{l=m}^{L-1} \lambda_l^{(3)} - \mu } \\
  && ~~~ = h_m \left( H_m \left( \xi_{p_m} \right) \right) - 0 + \left( \xi_N - h_m \left( H_m \left(\xi_{p_m} \right) \right) \right) - 0 - \xi_{N} = 0.
\end{eqnarray*}

The output $y^*$ of Algorithm \ref{alg:universal} and the Lagrange multiplier assignments satisfy the KKT conditions; $y^*$ therefore minimizes (\ref{eqn:minimizeSumSeparableConvexFunctions}), and the proof of Theorem \ref{thm:correctnessOfAlgorithm} is complete. \hspace*{\fill}~\QEDclosed

\subsection{Proof of Corollary \ref{cor:orderedConditionCorollary}}
The assignments in Algorithm \ref{alg:universal} are
\[
  y_m^* = H_m \left( \xi_{p_m} \right), m = \llbracket  1,L  \rrbracket .
\]
By hypothesis, $H_m \geq H_{m+1}, m \in \llbracket  1,L-1  \rrbracket $, and by Proposition \ref{prop:correctness}, $\xi_n \geq \xi_{n+1}, n \in \llbracket  1,N-1  \rrbracket $. These monotonicity properties imply
\[
  y_m^* = H_m \left( \xi_{p_m} \right) \geq H_{m+1} \left( \xi_{p_m} \right) \geq H_{m+1} \left( \xi_{p_{m+1}} \right) = y_{m+1}^*, m \in \llbracket  1,L-1  \rrbracket .
\]
\hspace*{\fill}~\QEDclosed

\subsection{Proof of Proposition \ref{prop:valueFunctionProperties}}:
Recall that here $K = L$. Define
\[
  \mathcal{L}(\alpha) := \left\{ y \in \mathbb{R}^L : y \mbox{ satisfies } (\ref{eqn:positivityBounded})-(\ref{eqn:equalityConstraint}) \right\}.
\]
$\mathcal{L}(\alpha)$ is convex, but may not be closed because the domains $(a_m, b_m)$ may not be closed. From the ascending constraints (\ref{eqn:ladderConstraint}) and (\ref{eqn:equalityConstraint}), it is clear that if $\alpha \succeq \tilde{\alpha}$ then $\mathcal{L}(\alpha) \subseteq \mathcal{L} \left( \tilde{\alpha} \right)$, and therefore $\mathcal{G}(\alpha) \geq \mathcal{G} \left( \tilde{\alpha} \right)$. The first statement is therefore proved. (Note that the conditions on $g_m, m\in\llbracket  1,L  \rrbracket $ stated in Section \ref{sec:problemDescription} are not necessary for this property).

To show convexity, consider $\alpha, \tilde{\alpha} \in \mathbb{R}_+^L$. Fix $\lambda \in (0,1)$. If either of $\mathcal{L}(\alpha)$ or $\mathcal{L} \left( \tilde{\alpha} \right)$ is empty, there is nothing to prove. We may therefore assume both are nonempty and therefore $\mathcal{G}(\alpha)$ and $\mathcal{G} \left( \tilde{\alpha} \right)$ are finite. For every $\varepsilon > 0$, there exist $y \in \mathcal{L}(\alpha)$ and $\tilde{y} \in \mathcal{L} \left( \tilde{\alpha} \right)$ satisfying $G(y) < \mathcal{G}(\alpha) + \varepsilon$ and $G \left(\tilde{y} \right) < \mathcal{G} \left( \tilde{\alpha} \right) + \varepsilon$. The linearity of the constraints implies $\lambda y + (1-\lambda) \tilde{y} \in \mathcal{L} \left( \lambda \alpha + (1 - \lambda) \tilde{\alpha} \right)$. The convexity of $G$ implies
\begin{eqnarray*}
  \mathcal{G} \left( \lambda \alpha + (1 - \lambda) \tilde{\alpha} \right) & \leq & G \left( \lambda y + (1-\lambda) \tilde{y} \right) \\
  & \leq & \lambda G(y) + (1-\lambda) G \left( \tilde{y} \right) \\
  & \leq & \lambda \mathcal{G}(\alpha) + (1 - \lambda) \mathcal{G} \left( \tilde{\alpha} \right) + \varepsilon.
\end{eqnarray*}
Since $\varepsilon$ is arbitrary, the convexity of $\mathcal{G}$ is established. \hspace*{\fill}~\QEDclosed

%\bibliographystyle{../sty/IEEEtran}
%{
%\bibliography{../sty/IEEEabrv,../wislBib/wisl}
%}

\bibliography{SeparableConvexOptimizationArxiv.bbl}

\end{document}